\documentclass[pre,tighten]{revtex4}

\usepackage{amssymb,amsmath}

\usepackage{graphicx}

\begin{document}
\title{Instabilities in a free granular fluid described by the Enskog equation}
\author{Vicente Garz\'{o}}
\address{Departamento de F\'{\i}sica, Universidad de Extremadura, E-06071
Badajoz, Spain}
\email[E-mail: ]{vicenteg@unex.es}
\begin{abstract}
A linear stability analysis of the hydrodynamic equations with
respect to the homogeneous cooling state is carried out to identify
the conditions for stability as functions of the wave vector, the
dissipation, and the density. In contrast to previous studies, this
description is based on the results derived from the Enskog equation
for inelastic hard spheres [V. Garz\'o and J. W. Dufty,
Phys. Rev. E {\bf 59}, 5895 (1999)], which takes into account the dependence of
the transport coefficients on dissipation. As expected, linear
stability shows two transversal (shear) modes and a longitudinal (``heat'')
mode to be unstable with respect to long enough wavelength excitations.
Comparison with previous results (which neglect the influence of
dissipation on transport) shows quantitative discrepancies for
strong dissipation.
\end{abstract}

\draft

\date{\today}
\maketitle

\section{Introduction}
\label{sec1}

A simple way of capturing the dynamics of granular media under rapid flow conditions
is through an idealized fluid of smooth, inelastic hard spheres. The inelasticity of
collisions is only accounted for by a (constant) coefficient of normal restitution $0<\alpha\leq 1$
that only affects the translational degrees of freedom of grains.
Despite the simplicity of the model, it has been widely used as a prototype to understand
some of the physical mechanisms involved in granular flows, especially those related to the
inelasticity of granular collisions. In particular,
one of the most characteristic features of granular fluids, as compared
with normal fluids, is the spontaneous formation of velocity vortices and
density clusters when evolving freely. This clustering instability can be
well described through a linear stability analysis of the
hydrodynamic equations and follows from the presence of a dissipation term
in the equation for the balance of energy.
An important feature of this instability is that it is confined to long
wavelengths (small wave numbers) and so it can be avoided for small enough systems.
First detected by Goldhirsch and
Zanetti \cite{GZ93} and McNamara \cite{M93}, the clustering problem has
attracted much attention in the past few years, especially from a
computational point of view \cite{BP04}.

In the case of a low-density gas, accurate predictions
for the unstable hydrodynamic modes have been made from the
(inelastic) Boltzmann equation \cite{BDKS98,DB03,DBZ04}. In particular,
a critical length $L_c$ is identified, so that the system
becomes unstable when its linear size
is larger than $L_c$. The dependence of $L_c$ on the coefficient of
restitution $\alpha$ predicted by kinetic theory compares quite well
with numerical results obtained by using the direct simulation Monte
Carlo method \cite{BRM98}.

For finite higher densities, these instabilities have been studied
by several authors using macroscopic or kinetic equations
\cite{GZ93,M93,varios}. A careful study of the dispersion relations
has been recently carried out by van Noije and Ernst \cite{NE00}
from the Enskog kinetic theory but neglecting any dependence of the pressure
and of the transport coefficients on inelasticity. Specifically,
they assume that the expressions for the hydrostatic pressure, the shear viscosity,
the bulk viscosity, and the thermal conductivity are the same as those for
the {\em elastic} gas \cite{CC70}, except that the temperature still depends explicitly
on time to account for the homogenous cooling.
However, given that the effect of inelasticity on dense fluid transport is
quite important in the undriven case \cite{GD99,GM02} (see for instance,
Fig.\ \ref{fig1rev} below) the assumptions
made in Ref.\ \cite{NE00} could be only justified for very small dissipation
(say for instance, $\alpha \approx 0.99$).

Although the predictions made by van Noije and Ernst
\cite{NE00} compares reasonably well with molecular dynamics simulation results,
it is worth to assess to what extent the previous results \cite{NE00} are indicative
of what happens when the improved expressions for the
{\em inelastic} transport coefficients
are considered \cite{GD99}. For this reason, in this paper I
revisit the (unstable) hydrodynamic-mode problem of a granular fluid described
by the {\em inelastic} Enskog kinetic theory \cite{BDS97}.
In spite of the explicit knowledge of the Enskog transport coefficients \cite{GD99},
I am not aware of any previous solution of the linearized hydrodynamic
equations for a moderately
dense granular gas.

The Enskog kinetic equation can be considered as the extension of
the Boltzmann equation to finite densities. As happens for elastic
collisions, the inelastic Enskog equation provides a semiquantitative
description of the hard sphere system that neglects the effects of
correlations between the velocities of the two particles that are
about to collide (molecular chaos assumption). The Enskog approximation
is expected to be valid for short times since as the system evolves corrections
to the Enskog equation due to multiparticle collisions, including
recollision events (``ring'' collisions) should be incorporated. The latter
are expected to be stronger for fluids with inelastic collisions where the
colliding pairs tend to be more focused. Therefore,
some deviations from molecular chaos have been observed in molecular
dynamics (MD) simulations \cite{MD} of granular fluids as the density
increases. Although the existence of these correlations restricts
the range of validity of the Enskog equation, the latter
can be still considered as a good approximation especially at
the level
of macroscopic properties (such as transport coefficients).
In particular, the Enskog results
presents quite a good agreement with MD simulations and even with real
experiments. In the case of computer simulations, comparison between the
Enskog theory and MD simulations in the case of the self-diffusion coefficient
\cite{LBD02} and kinetic temperatures in a granular mixture \cite{DHGD02} have
shown good agreement for all $\alpha$ at $n^*\equiv n\sigma^3\leq 0.25$ and for
all densities at $\alpha \geq 0.9$. Here, $n$ is the number density and $\sigma$
is the diameter of spheres. More recent agreement has been found in the case
of granular mixtures under shear flow \cite{AL03}. The Enskog transport coefficients
\cite{GD99} have also been tested against real NMR experiments of a system of mustard seeds
vibrated vertically \cite{YHCMW02,HYCMW04}. The averaged value of the coefficient of restitution
of the grains used in this experiment is $\alpha=0.87$, which lies outside of the
quasielastic limit ($\alpha \approx 0.99$). Comparison between theory and experiments
(see for instance, Figs.\ 10--13 of Ref.\ \cite{HYCMW04}) shows that
the Enskog kinetic theory \cite{GD99} successfully models the density and granular
temperature profiles away from the vibrating container bottom and
quantitatively explains the temperature inversion observed in experiments.
All these results clearly show the applicability of the Enskog theory for
densities outside the Boltzmann limit ($n^* \to 0$) and values of dissipation beyond
the quasielastic limit. In this context, one can conclude that the Enskog equation
provides a unique
basis for the description of dynamics across a wide range of densities,
length scales, and degrees of dissipation. No other theory with such
generality exists.

The explicit knowledge of the Navier-Stokes transport coefficients as well as
of the cooling rate for inelastic hard spheres \cite{GD99} allows one to
solve the linearized hydrodynamic equations around the homogeneous cooling
state (HCS) and identify the conditions for stability as functions of the wave
vector, the dissipation, and the density. In the low-density limit, previous results
derived from the Boltzmann equation are recovered \cite{BDKS98}.
Linear stability analysis
shows two transversal (shear) modes and a longitudinal (``heat'') mode to be unstable
with respect to long wavelength excitations. The corresponding critical values for the
shear and heat modes are also determined, showing that the clustering instability
is mainly driven by the transversal shear mode, except for quite large dissipation.
As expected, these results agree qualitatively well
with those previously derived in Ref.\ \cite{NE00}. On the other hand, at a
quantitative level, the comparison carried out here shows
significant differences between both descriptions as the collisions become
more inelastic.

The plan of the paper is as follows. In Sec.\ \ref{sec2}, the basis
of the hydrodynamic equations for a dense gas as derived from the (inelastic)
Enskog equation is described. The explicit dependence of the transport coefficients
and the cooling rate on dissipation is also illustrated for some values of density to show
that the influence of inelasticity on transport is in general quite significant. Section
\ref{sec3} is devoted to the linear stability analysis around the HCS
and presents the main results of this paper. The paper
is closed in Sec.\ \ref{sec4} with some concluding remarks.

\section{Hydrodynamic description}
\label{sec2}

We consider a granular fluid composed of smooth inelastic hard
spheres of mass $m$ and diameter $\sigma$. Collisions are
characterized by a (constant) coefficient of normal restitution
$0<\alpha\leq 1$. At a kinetic level, all the relevant information
on the system is given through the one-particle velocity
distribution function, which is assumed to obey the (inelastic) Enskog equation \cite{BDS97}.
From it one can easily get the (macroscopic) hydrodynamic equations
for the number density $n({\bf r}, t)$, the flow velocity ${\bf
u}({\bf r}, t)$, and the local temperature $T({\bf r}, t)$ \cite{GD99}:
\begin{equation}
D_{t}n+n \nabla \cdot {\bf u}=0\;,  \label{1}
\end{equation}
\begin{equation}
\rho D_{t}{\bf u}+\nabla {\sf P}=0\;,  \label{2}
\end{equation}
\begin{equation}
D_{t}T+\frac{2}{3n}\left(\nabla \cdot {\bf q}+{\sf P}:\nabla {\bf
u}\right)=-\zeta T.
\label{3}
\end{equation}
In the above equations, $D_{t}=\partial _{t}+{\bf u}\cdot \nabla $
is the material derivative,  $\rho=nm$ is the mass density, ${\sf
P}$ is the pressure tensor, ${\bf q}$ is the heat flux, and $\zeta$
is the cooling rate due to the energy dissipated in collisions. The
practical usefulness of the balance equations (\ref{1})--(\ref{3}) is limited
unless the fluxes and the cooling rate are further specified in
terms of the hydrodynamic fields and their gradients. The detailed form of
the constitutive equations and the transport coefficients appearing in
them have been obtained by applying the Chapman-Enskog method \cite{CC70}
to the Enskog equation. To first order in the gradients, the corresponding
constitutive equations are  \cite{GD99}
\begin{equation}
P_{ij}=p\delta _{ij}-\eta \left( \nabla _{j}u_{i }+\nabla _{i
}u_{j}-\frac{2}{3}\delta _{ij} {\bf \nabla \cdot u}\right)-\gamma
\delta_{ij} \nabla \cdot {\bf u},
 \label{4}
\end{equation}
\begin{equation}
{\bf q}=-\kappa \nabla T-\mu \nabla n,
\label{5}
\end{equation}
\begin{equation}
\label{6}
\zeta=\zeta_0+\zeta_1 \nabla \cdot {\bf u}.
\end{equation}
Here, $p$ is the hydrostatic pressure, $\eta$ is the shear
viscosity, $\gamma$ is the bulk viscosity, $\kappa$ is the thermal
conductivity, and $\mu$ is a new transport coefficient not present
in the elastic case. The expressions for the pressure, the transport
coefficients and the cooling rate can be written in the forms
\begin{equation}
\label{7}
p=nT p^*(\alpha, \phi), \quad \eta=\eta_0 \eta^*(\alpha, \phi),
\quad \gamma=\eta_0 \gamma^*(\alpha, \phi),
\end{equation}
\begin{equation}
\label{8}
\kappa=\kappa_0 \kappa^*(\alpha, \phi), \quad
\mu=\frac{T\kappa_0}{n} \mu^*(\alpha, \phi), \quad
\zeta_0=\frac{nT}{\eta_0} \zeta_0^*(\alpha, \phi),
\end{equation}
where $\eta_0=5(mT)^{1/2}/16\sigma^2 \pi^{1/2}$ and
$\kappa_0=15\eta_0/4m$ are the low-density values of the shear
viscosity and the thermal conductivity in the elastic limit,
respectively. The quantities $p^*$, $\eta^*$, $\gamma^*$,
$\kappa^*$, $\mu^*$, $\zeta_0^*$, and $\zeta_1$ are dimensionless
functions of the coefficient of restitution $\alpha$ and the solid
volume fraction $\phi=\pi n \sigma^3/6$. Their explicit expressions
are given in Appendix \ref{appA}, and more details can be found
in Ref.\ \cite{GD99}. For elastic collisions ($\alpha=1$), $\mu^*(1,\phi)$,
$\zeta_0(1,\phi)^*$, and $\zeta_1(1,\phi)$ vanish, while the expressions of $\eta^*(1,\phi)$,
$\gamma^*(1,\phi)$, and $\kappa^*(1,\phi)$ coincide with those obtained for
a dense gas of elastic hard spheres \cite{CC70}. In the low-density
limit ($\phi=0$), $\gamma^*(\alpha,0)=\zeta_1(\alpha,0)=0$ and the results
derived for a dilute granular gas are recovered \cite{BDKS98}.
As pointed out before, the new transport coefficient $\mu$ is not present
for elastic collisions and may play an important role to accurately describe
some situations of real granular materials, such as a ``temperature
inversion'' observed in vibrofludized systems \cite{HYCMW04}.

\begin{figure}
\includegraphics[width=0.5 \columnwidth,angle=0]{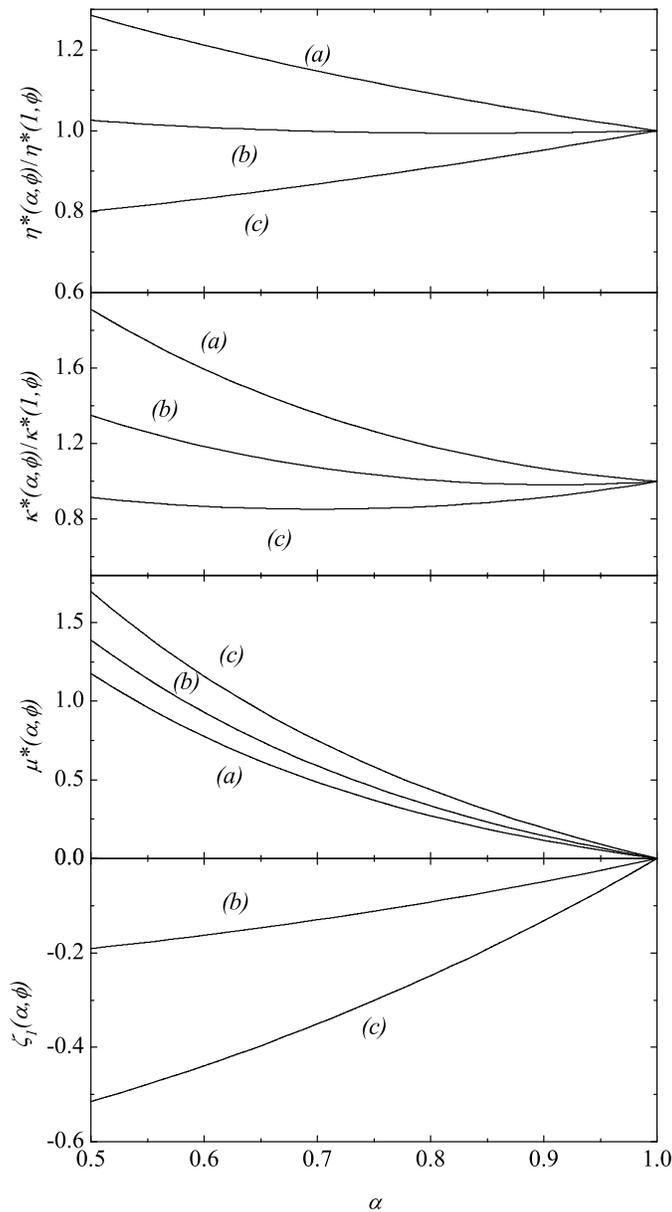}
\caption{Plot of $\eta^*(\alpha,\phi)/\eta^*(1,\phi)$, $\kappa^*(\alpha,\phi)/\kappa^*(1,\phi)$,
$\mu^*(\alpha,\phi)$, and $\zeta_1(\alpha,\phi)$ versus the coefficient of restitution $\alpha$ for
three different values of the solid volume fraction $\phi$: (a) $\phi=0$, (b) $\phi=0.1$,
and (c) $\phi=0.2$. Note that $\zeta_1=0$ at $\phi=0$.
\label{fig1rev}}
\end{figure}
The reduced quantities $\eta^*(\alpha,\phi)/\eta^*(1,\phi)$, $\kappa^*(\alpha,\phi)/\kappa^*(1,\phi)$,
$\mu^*(\alpha,\phi)$, and $\zeta_1(\alpha,\phi)$ are plotted in Fig.\ \ref{fig1rev}
as functions of the coefficient of restitution for three different values of
the solid volume fraction $\phi$.
As said in the Introduction, these
quantities, along with the pressure, were assumed to be the same as in the elastic case in the
stability analysis carried out in Ref.\ \cite{NE00}, i.e.,
$p^*(\alpha,\phi)\to p^*(1,\phi)$,
$\eta^*(\alpha,\phi)\to \eta^*(1,\phi)$, $\kappa^*(\alpha,\phi) \to \kappa^*(1,\phi)$, and
$\mu^*(\alpha,\phi)=\zeta_1(\alpha,\phi)\to 0$.
Figure \ref{fig1rev} shows that in general
the influence of dissipation on the transport coefficients and the cooling rate is quite
significant and so their functional form differs appreciably from their elastic form.
This means that the predictions made in Ref.\ \cite{NE00} might
quantitatively differ from those obtained here as the rate of dissipation increases. This will be
confirmed later. We also see that, for a given value of $\alpha$,
$\eta^*(\alpha,\phi)/\eta^*(1,\phi)$ and $\kappa^*(\alpha,\phi)/\kappa^*(1,\phi)$
decrease as the density increases, while the opposite happens in the cases of $\mu^*(\alpha,\phi)$ and $|\zeta_1(\alpha,\phi)|$.

When the expressions of the pressure tensor, the heat flux and the cooling rate are substituted
into the balance equations (\ref{1})--(\ref{3}) one gets the corresponding Navier-Stokes (closed)
hydrodynamic equations for $n$, ${\bf u}$ and $T$. They are given by
\begin{equation}
\label{2.1}
D_tn+n\nabla \cdot {\bf u}=0,
\end{equation}
\begin{equation}
\label{2.2}
D_t u_i+(nm)^{-1}\nabla_i p=(nm)^{-1}\nabla_j\left[\eta \left(\nabla_iu_j+\nabla_ju_i
-\frac{2}{3}
\delta_{ij}\nabla \cdot {\bf u}\right)+\gamma \delta_{ij}\nabla \cdot {\bf u}\right],
\end{equation}
\begin{eqnarray}
\label{2.3}
\left(D_t+\zeta_0\right)T&+&\frac{2}{3n}p\nabla \cdot {\bf u}=\frac{2}{3n}\nabla \cdot \left
(\kappa \nabla T+\mu \nabla n\right)\nonumber\\
& & +\frac{2}{3n}\left[\eta \left(\nabla_iu_j+\nabla_ju_i-\frac{2}{3}
\delta_{ij}\nabla \cdot {\bf u}\right)
+\gamma \delta_{ij}\nabla \cdot {\bf u}\right]
-
T\zeta_1  \nabla\cdot {\bf u}.
\end{eqnarray}
Note that consistency would require to consider up to second order in the gradients in the expression (\ref{6})
for the cooling rate, since this is the order of the terms in Eq.\ (\ref{2.3}) coming from
the pressure tensor and the heat flux.
However, it has been shown for a dilute gas that the contributions from the cooling rate
of second order are negligible as compared with the corresponding contributions from Eqs.\ (\ref{4})--(\ref{6})
\cite{BDKS98}. It is assumed here that the same holds in the dense case.

The form of the Navier-Stokes equations (\ref{2.1})--(\ref{2.3}) is the same as for a normal fluid, except for
the presence of the contributions to the cooling rate $\zeta_0$ and $\zeta_1$ and the new transport coefficient
$\mu$ in the energy balance equation. Of course, as Fig.\ \ref{fig1rev} clearly illustrates, the values
of the transport coefficients are quite different, depending on the value of the coefficient of restitution $\alpha$.

\section{Linear stability analysis}
\label{sec3}

The hydrodynamic equations (\ref{1})--(\ref{3}) admit a simple solution
which corresponds to the so-called homogeneous cooling state (HCS). It describes a
uniform state with vanishing flow field and a temperature decreasing
monotonically in time, i.e.,
\begin{equation}
\label{3.1}
T(t)=\frac{T(0)}{\left(1+\zeta_{0}(0)t/2\right)^{2}}.
\end{equation}
Nevertheless, computer simulations \cite{GZ93,NY94,BRC99} have shown that the HCS is unstable
with respect to
long enough wavelength perturbations. To analyze this problem, it is adequate to perform a stability
analysis of the nonlinear hydrodynamic equations (\ref{2.1})--(\ref{2.3}) with respect to
the homogeneous state for small initial excitations. The linearization of
Eqs.\ (\ref{2.1})--(\ref{2.3}) about
the homogenous solution
yields partial differential equations with coefficients that are
independent of space but depend on time since the reference (homogeneous)
state is cooling. This time dependence can be eliminated through a change in
the time and space variables and a scaling of the hydrodynamic
fields.

We assume that the deviations $\delta y_{\alpha}({\bf
r},t)=y_{\alpha}({\bf r},t)-y_{H \alpha}(t)$ are small, where,
$\delta y_{\alpha}({\bf r},t)$ denotes the deviation of $\{n, {\bf
u}, T,\}$ from their values in the homogeneous state, the latter being denoted by
the subscript $H$. The quantities in the HCS verify
\begin{equation}
\label{9}
\nabla n_H=\nabla T_H=0, \quad {\bf u}_H={\bf 0}, \quad \partial_t
\ln T_H=-\zeta_{0H}.
\end{equation}
To recover the results found in the dilute gas case when $\phi \to 0$,
I consider the same time and space variables as those used in \cite{BDKS98}, namely,
\begin{equation}
\label{3.2}
\tau=\frac{1}{2}\int_0^t\;\nu_{H}(t')dt', \quad {\boldsymbol{\ell}}=\frac{1}{2}
\frac{\nu_{H}(t)}{v_{H}(t)}{\bf r},
\end{equation}
where $\nu_H(t)=\frac{16}{5}n_H\sigma^2\pi^{1/2}v_H(t)$ is an effective
collision frequency and $v_H(t)=\sqrt{T_H(t)/m}$. Note that $\nu_H(t)=n_HT_H/\eta_{H}(1,0)$ is
an effective collision frequency associated with the elastic ($\alpha=1$)
shear viscosity of a dilute gas ($\phi=0$).
The dimensionless time scale $\tau$ is the time integral of the average collision
frequency and thus is a measure of the average number of collisions per particle
in the time interval between $0$ and $t$. The unit length $v_H(t)/\nu_H(t)$ introduced in
the second equality of (\ref{3.2}) is proportional to the time-independent mean free path of gas
particles.

A set of Fourier transformed dimensionless variables are then
introduced by
\begin{equation}
\label{10}
\rho_{{\bf k}}(\tau)=\frac{\delta n_{{\bf k}}(\tau)}{n_{H}}, \quad
{\bf w}_{{\bf k}}(\tau)=\frac{\delta {\bf u}_{{\bf
k}}(\tau)}{v_H(\tau)},\quad \theta_{{\bf k}}(\tau)=\frac{\delta
T_{{\bf k}}(\tau)}{T_{H}(\tau)},
\end{equation}
where $\delta y_{{\bf k}\alpha}\equiv \{\delta n_{{\bf k}},{\bf
w}_{{\bf k}}(\tau), \theta_{{\bf k}}(\tau)\}$ is defined as
\begin{equation}
\label{11}
\delta y_{{\bf k}\alpha}(\tau)=\int d{\boldsymbol {\ell}}\;
e^{-i{\bf k}\cdot {\boldsymbol {\ell}}}\delta y_{\alpha}
({\boldsymbol {\ell}},\tau).
\end{equation}
Note that in Eq.\ (\ref{11}) the wave vector ${\bf k}$ is dimensionless.
In terms of the above variables, the transverse velocity components
${\bf w}_{{\bf k}\perp}={\bf w}_{{\bf k}}-({\bf w}_{{\bf k}}\cdot
\widehat{{\bf k}})\widehat{{\bf k}}$ (orthogonal to the wave vector ${\bf k}$)
decouple from the other three modes and hence can be obtained more
easily. Their evolution equation is
\begin{equation}
\label{16}
\left(\frac{\partial}{\partial \tau}-\zeta_0^*+\frac{1}{2}\eta^*
k^2\right){\bf w}_{{\bf k}\perp}=0,
\end{equation}
where it is understood that $\zeta_0^*$ and $\eta^*$ are evaluated in the HCS.
The solution to Eq.\ (\ref{16}) is
\begin{equation}
\label{16.1}
{\bf w}_{{\bf k}\perp}({\bf k}, \tau)={\bf w}_{{\bf k}\perp}(0)\exp[s_{\perp}(k)\tau],
\end{equation}
where
\begin{equation}
\label{16.2}
s_{\perp}(k)=\zeta_0^*-\frac{1}{2}\eta^* k^2.
\end{equation}
This identifies two shear (transversal) modes analogous to the
elastic ones. According to Eq.\ (\ref{16.2}), there exists
a critical wave number $k_{s}$ given by
\begin{equation}
\label{16.3}
k_{s}=\left(\frac{2\zeta_0^*}{\eta^*}\right)^{1/2}.
\end{equation}
This critical value separates two regimes: shear modes with $k> k_{s}$ always decay
while those with $k< k_{s}$ grow exponentially.

The remaining (longitudinal) modes correspond to $\rho_{{\bf k}}$, $\theta_{{\bf k}}$, and
the longitudinal velocity component of the velocity field, $w_{{\bf k}||}={\bf w}_{{\bf
k}}\cdot \widehat{{\bf k}}$ (parallel to ${\bf k}$). These modes are coupled and obey the equation
\begin{equation}
\frac{\partial \delta y_{{\bf k}\alpha }(\tau )}{\partial \tau }=M_{\alpha \beta}
 \delta y_{{\bf k}\beta }(\tau ),
\label{17}
\end{equation}
where $\delta y_{{\bf k}\alpha }(\tau )$ denotes now the set  $\left\{ \rho _{{\bf k}},\theta _{{\bf k}},
 w_{{\bf k}||}\right\}$ and ${\sf M}$ is the square matrix
\begin{widetext}
\begin{equation}
{\sf M}=\left(
\begin{array}{ccc}
0 & 0 & -i k \\
-2\zeta_0^*g-\frac{5}{4}\mu^*k^2&-\zeta_0^*-\frac{5}{4}\kappa^*k^2 & -\frac{2}{3}ik(p^*+\frac{3}{2}\zeta_1)\\
-ikp^*C_\rho & -ikp^* &\zeta_0^{*}-\frac{2}{3}\eta^*k^2-\frac{1}{2}\gamma^*k^2
\end{array}
\right).   \label{18}
\end{equation}
\end{widetext}
As before, it is understood that $p^*$, $\eta^*$, $\gamma^*$, $\kappa^*$, $\mu^*$,
$\zeta_0^*$, and $\zeta_1$ are evaluated in the HCS. In addition, the quantities $g(\phi)$ and
$C_\rho(\alpha,\phi)$ are given by
\begin{equation}
\label{19}
g(\phi)=1+\phi\frac{\partial}{\partial \phi}\ln \chi(\phi),
\end{equation}
\begin{eqnarray}
\label{20}
C_\rho(\alpha,\phi)&=&1+g(\phi)\frac{p^*(\alpha,\phi)-1}{p^*(\alpha,\phi)}\nonumber\\
&=&1+g(\phi)-\frac{g(\phi)}{1+2(1+\alpha)\phi \chi(\phi)},
\end{eqnarray}
where in the last equality use has been made of the explicit expression of $p^*$
given by Eq.\ (\ref{a1}). In Eqs.\ (\ref{19}) and (\ref{20}), $\chi(\phi)$ is the pair correlation
function at contact.  In kinetic theory calculations,
the value of $\chi$ for the pre-collisional
distribution is used, which is well approximated by local
equilibrium. There are nonequilibrium corrections
that can be calculated from the ring collision
operator \cite{L01}. However, these corrections are very hard to
calculate and so for simplicity I take here
the Carnahan-Starling approximation \cite{CS69}:
\begin{equation}
\label{21}
\chi(\phi)=\frac{2-\phi}{2(1-\phi)^3}.
\end{equation}
In the limit $\phi\to 0$, $p^*=g=C_\rho=1$, $\gamma^*=\zeta_1=0$ and Eqs.\
(\ref{16})--(\ref{18}) reduce to those previously derived for a
dilute gas \cite{BDKS98}.
\begin{figure}
\includegraphics[width=0.5 \columnwidth,angle=0]{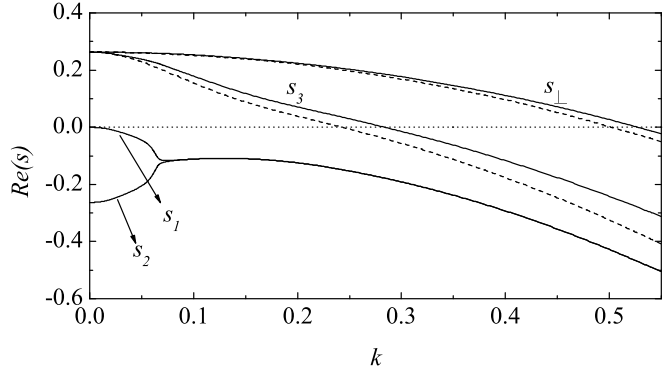}
\caption{Dispersion relations for a granular fluid with $\alpha=0.8$ and
$\phi=0.2$. From top to bottom the curves correspond to the two degenerate shear (transversal) modes and
the remaining three longitudinal modes. The dashed lines correspond to the results obtained
for the shear ($s_{\perp}$) and heat ($s_3$) modes from the approximations made in Ref.\ \cite{NE00}.
Only the real parts of the eigenvalues is plotted.
\label{fig2bis}}
\end{figure}

The longitudinal three modes have the form $\exp[s_n(k) \tau]$ for $n=1,2,3$, where
$s_n(k)$ are the eigenvalues of the matrix ${\sf M}$, namely,
they are the solutions of the cubic equation
\begin{eqnarray}
\label{22}
& &
s^3+\frac{5}{4}\left(\kappa^*+\frac{2}{5}\gamma^*+\frac{8}{15}\eta^*\right)k^2
s^2+\left\{ k^4
\kappa^* \left(\frac{5}{6}\eta^*+\frac{5}{8}\gamma^*\right)
+k^2\left[p^*
C_\rho+\frac{2}{3}\eta^*\zeta_0^*+\frac{1}{2}\gamma^*\zeta_0^*\right.\right.\nonumber\\
& & \left.\left.+\frac{1}{3}p^*
(2p^*+3\zeta_1)-\frac{5}{4}\kappa^*\zeta_0^*\right]-\zeta_0^{*2}\right\}s
+p^*\left[\frac{5}{4}(\kappa^*C_\rho-\mu^*)k^2+\zeta_0^*(C_\rho-2
g)\right]k^2=0.
\end{eqnarray}
As happens for a dilute gas \cite{BDKS98}, for given values of $\alpha$ and $\phi$
the analysis of Eq.\ (\ref{22}) shows that
at very small $k$ all modes are real, while at larger $k$ two modes
become a complex conjugate pair of propagating modes. Thus, the
physical meaning of the longitudinal modes is different from that in
the elastic fluids, even when $\alpha \to 1$ \cite{BDKS98}.

\begin{figure}
\includegraphics[width=0.5 \columnwidth,angle=0]{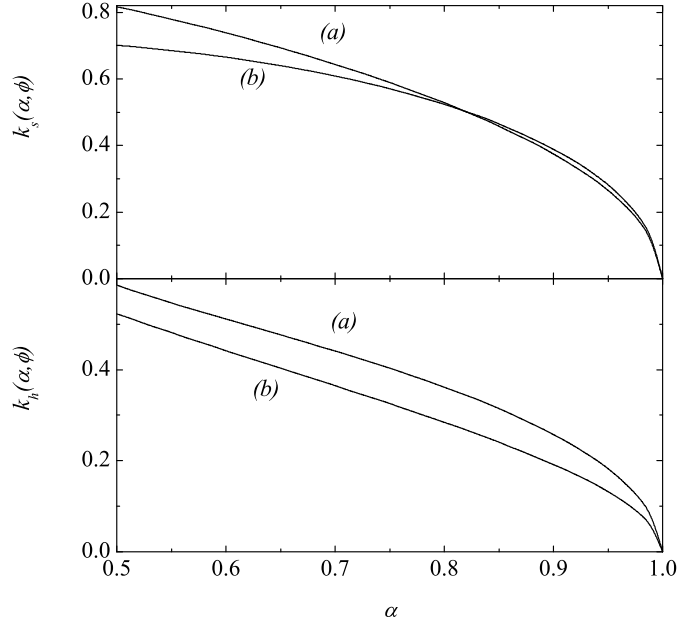}
\caption{Plot of the the critical wave numbers $k_s(\alpha,\phi)$ and $k_h(\alpha,\phi)$
as functions of the coefficient of
restitution $\alpha$ for two values of the solid volume fraction $\phi$: (a) $\phi=0$
and (b) $\phi=0.2$.
\label{fig4rev}}
\end{figure}

The solution to Eq.\ (\ref{22}) can be obtained for small $k$ by a
perturbation expansion as
\begin{equation}
\label{23}
s_{n}(k)=s_{n}^{(0)}+k s_n^{(1)}+k^2s_{n}^{(2)}+\cdots.
\end{equation}
Substituting this expansion into Eq.\ (\ref{22}) yields $s_1^{(0)}=0$, $s_2^{(0)}=-\zeta_0^*$,
$s_3^{(0)}=\zeta_0^*$,
\begin{equation}
\label{25}
s_n^{(1)}=0,
\end{equation}
\begin{equation}
\label{26}
s_{1}^{(2)}=\frac{p^*}{\zeta_0^*}\left(C_\rho-2g\right),
\end{equation}
\begin{equation}
\label{27}
s_{2}^{(2)}=p^*\frac{2p^*+3(\zeta_1+2g)}{6\zeta_0^*}-\frac{5}{4}\kappa^*,
\end{equation}
\begin{equation}
\label{28}
s_{3}^{(2)}=-p^*\frac{2p^*+3(\zeta_1-2g)+6C_\rho}{6\zeta_0^*}
-\frac{2}{3}\eta^*-\frac{1}{2}\gamma^*.
\end{equation}
In the case of a dilute gas ($\phi\to 0$), the eigenvalues $s_n(k)$ behave as
\begin{equation}
\label{28.1}
s_1(k)\to -\frac{1}{\zeta_0^*}k^2,
\end{equation}
\begin{equation}
\label{28.2}
s_2(k)\to -\zeta_0^*+\left(\frac{4}{3\zeta_0^*}-\frac{5}{4}\kappa^*\right)k^2,
\end{equation}
\begin{equation}
\label{28.3}
s_3(k)\to \zeta_0^*-\left(\frac{1}{3\zeta_0^*}+\frac{2}{3}\eta^*\right)k^2.
\end{equation}
Since the Navier-Stokes hydrodynamic equations  are valid to second order in $k$, the
solutions (\ref{26})--(\ref{28}) are relevant to the same order.

As said in the Introduction, although the limitations of the Enskog theory
are greater than for elastic systems, comparison with
MD simulations \cite{LBD02,DHGD02} indicate that
it is still accurate for $\phi$ up to about 0.15 and for $\alpha$ greater
than about 0.5. For higher densities the $\alpha$ range is more limited, but
even then it captures the relevant qualitative features. For this reason,
to illustrate the influence of both density and dissipation on instabilities,
densities in the interval $0\leq \phi \leq 0.2$ for
$0.5\leq \alpha \leq 1$ will be considered.

The dispersion relations $s_{n}(k)$ for a fluid with $\alpha=0.8$
and $\phi=0.2$, as obtained from Eq.\ (\ref{16.3}) and the solutions of the cubic
equation (\ref{22}), are plotted in Fig.\ \ref{fig2bis}. Only the real part
(propagating modes) of the solutions to Eq.\ (\ref{22}) is represented.
For comparison, the results derived for the shear ($s_{\perp}$)
and the longitudinal ``heat'' ($s_3$) modes
from the approximations made
by van Noije and Ernst \cite{NE00} are also plotted. These curves can be formally
obtained from the results derived in this paper
when one takes $\mu^*=\zeta_1=0$, and $p^*$, $\eta^*$, $\gamma^*$, and
$\kappa^*$ are replaced by their values in the elastic limit
[Eqs.\ (\ref{a17})--(\ref{a19})].
We observe that the agreement between both sets of results is
in general good, the heat mode showing more quantitative discrepancies.
Figure \ref{fig2bis} also shows that the
heat mode is unstable for $k<k_{h}$, where $k_{h}$ can be obtained from
Eq.\ (\ref{22}) when $s=0$. The result is
\begin{equation}
\label{29}
k_{h}=\sqrt{\frac{4\zeta_0^*(2g-C_\rho)}{5(\kappa^*C_\rho-\mu^*)}}.
\end{equation}
\begin{figure}
\includegraphics[width=0.5 \columnwidth,angle=0]{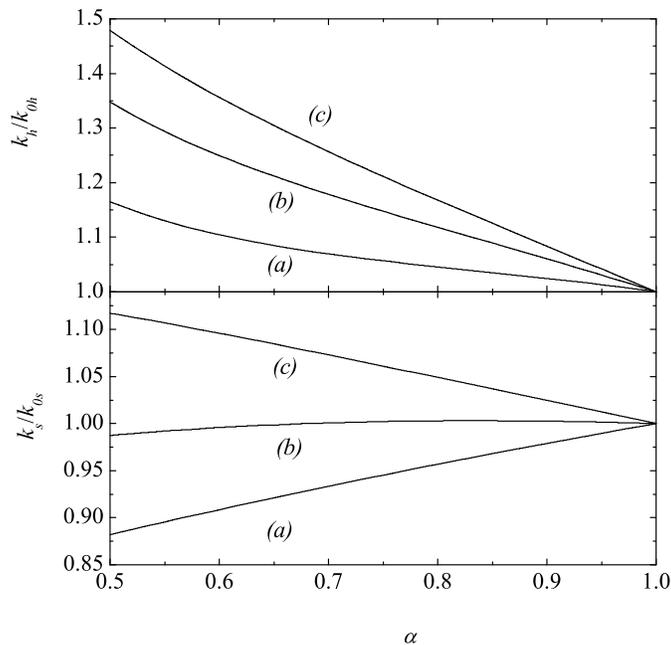}
\caption{Plot of the ratios $k_s/k_{0s}$ and $k_h/k_{0h}$ as functions of
the coefficient of
restitution $\alpha$ for three values of the solid volume fraction $\phi$: (a) $\phi=0$,
(b) $\phi=0.1$
and (c) $\phi=0.2$. Here, $k_{0s}$ and $k_{0h}$ are the critical wave numbers
obtained from the approximations made in Ref.\ \cite{NE00}.
\label{fig5rev}}
\end{figure}
\begin{figure}
\includegraphics[width=0.5 \columnwidth,angle=0]{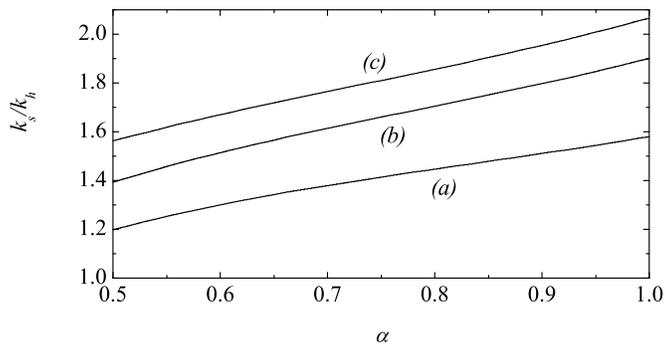}
\caption{Ratio $k_s/k_h$ versus the coefficient of restitution $\alpha$
for three
values of the solid volume fraction $\phi$: (a) $\phi=0$,
(b) $\phi=0.1$, and (c) $\phi=0.2$.
\label{fig6rev}}
\end{figure}
The dependence of the critical values $k_{s}$ and $k_{h}$ on dissipation is illustrated
in Fig\ \ref{fig4rev} for two values of $\phi$. For a given
value of the coefficient of restitution $\alpha$, in general the corresponding critical values decrease
with increasing density. However, there is a small region of values of $\alpha \gtrsim 0.82$ where
the opposite happens in the case of $k_s$.  All the above trends are
also captured by the results obtained in Ref.\ \cite{NE00}, although quantitative
discrepancies between both descriptions appear as the dissipation increases.
To illustrate such differences, the ratios  $k_s/k_{0s}$ and $k_h/k_{0h}$
are plotted versus $\alpha$ in Fig.\ \ref{fig5rev} for different values of $\phi$. Here,
$k_{0s}$ and $k_{0h}$ are the critical wave numbers
obtained from the approximations made in Ref.\ \cite{NE00}.
Significant differences between both analyses are clearly shown in Fig.\ \ref{fig5rev},
especially for strong dissipation and moderate densities.
Thus, for instance, for $\phi=0.2$ and $\alpha=0.8$ the discrepancies
between both approaches for $k_{s}$ and $k_{h}$ are about $5\%$ and $17\%$,
respectively, while for $\phi=0.2$ and $\alpha=0.5$ the discrepancies are about
 $12\%$ and $48\%$, respectively.
A more significant qualitative difference between our results and those
obtained in Ref.\ \cite{NE00} appears in the case of the ratio $k_{s}/k_h$. This quantity
measures the separation between both critical modes. According to Eqs.\ (\ref{16.3}) and
(\ref{29}), this ratio is independent of $\alpha$ when one neglects the influence of dissipation
on the pressure and the transport coefficients. However, the present results predict a
complex dependence of $k_{s}/k_h$ on $\alpha$. To illustrate it, the ratio $k_{s}/k_h$
is plotted versus the coefficient of restitution $\alpha$ in
Fig.\ \ref{fig6rev} for different values of density. It is apparent that in general
both critical values $k_{s}$ and $k_h$ are well separated, especially for small inelasticity.
The results also show that the influence of dissipation on the
ratio $k_{s}/k_h$ is less significant as the system becomes denser.
In addition, for a given value of $\phi$, there exists
a value of the coefficient of restitution $\alpha_0(\phi)$ for which $k_h>k_s$
for values of $\alpha<\alpha_0$. The dependence of $\alpha_0$ on the solid volume fraction $\phi$
is plotted in Fig.\ \ref{fig9bis}. It is apparent that
the value of $\alpha_0$ decreases with increasing
density. However, given that
the values of $\alpha_0$ are quite small, one can conclude that in practice
($\alpha \gtrsim 0.375$) the
instability of the system is driven by the
transversal shear mode since $k_s>k_h$ for $\alpha > \alpha_0(\phi)$.

According to these results, for not quite
extreme values of dissipation ($\alpha \gtrsim 0.375$),
three different regions can be identified. For $k>k_{s}$ all modes are negative and the system
is linearly stable with respect to initial perturbations with wave number in
this range (short wavelength region). For $k_{h}<k<k_{s}$, the
shear mode is unstable while the heat mode is linearly stable.
In this range the density (coupled to the heat mode) is also stable and
so, density inhomogeneities can only be created due to the nonlinear
coupling with the unstable shear mode \cite{BRC99}. Finally, if
$k<k_{s}$ first vortices and then clusters are developed and the
final state of the system is strongly inhomogeneous. A more detailed
analysis of the evolution of the granular gas can be found in Ref.\
\cite{BP04}.
\begin{figure}
\includegraphics[width=0.5 \columnwidth,angle=0]{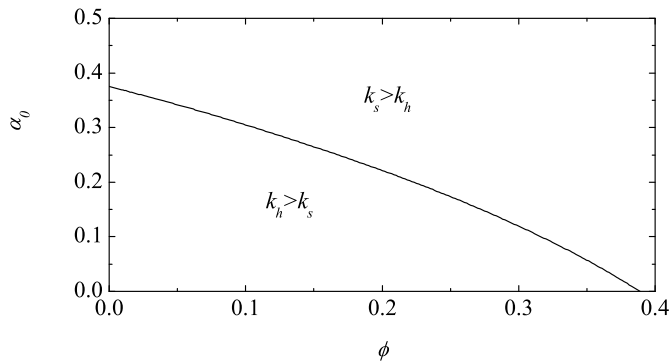}
\caption{Dependence of $\alpha_0$ on the solid volume fraction $\phi$. Points above (below) the
curve correspond to systems where the instability is driven by the shear (heat) mode.
\label{fig9bis}}
\end{figure}

\begin{figure}
\includegraphics[width=0.5 \columnwidth,angle=0]{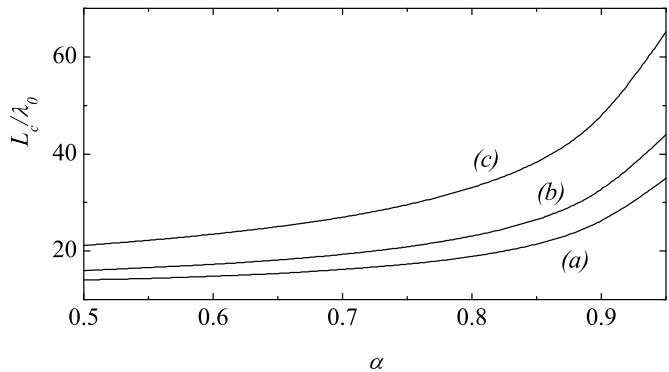}
\caption{The critical size $L_c$ in units of the mean free path $\lambda_0$ as a function
of the coefficient of restitution $\alpha$ for
three different values of the solid volume fraction $\phi$: (a) $\phi=0$, (b) $\phi=0.1$,
and (c) $\phi=0.2$. In each case, the system is linearly stable for points below the
corresponding curve.
\label{fig9}}
\end{figure}

In a system with periodic boundary conditions, the smallest allowed wave
number is $2\pi/L$, where $L$ is the largest system length.
Hence, for given values of inelasticity and density, we can identify
a critical length $L_c$ so that the system becomes unstable when
$L>L_c$. The value of $L_c$ is determined by equating
\begin{equation}
\label{30}
\frac{2\pi}{L_c^*}=\text{max}\{k_{s},k_h\},\quad L_c^*=\frac{\nu_H}{2v_H}L_c.
\end{equation}
In Fig.\ \ref{fig9} we show
$L_c/\lambda_0$ as a function of
$\alpha$ for different values of the solid volume fraction $\phi$.
Here, $\lambda_0=(\sqrt{2}\pi n\sigma^2 \chi)^{-1}=
\left(5\sqrt{2\pi}\chi/16\right)^{-1}v_H/\nu_H$ is the mean free
path of a hard-sphere dense gas. In all of these systems, $k_s>k_h$ and so
\begin{equation}
\label{26}
L_c=\frac{5}{4}\pi\sqrt{\pi}\chi\sqrt{\frac{\eta^*}{\zeta_0^*}}\lambda_0.
\end{equation}
For a given value of $\alpha$, we see that the critical size
(in units of the mean free path) increases with density. As a
consequence, larger systems are required to observe the shearing instability
as the fluid becomes denser.

\section{Concluding remarks}
\label{sec4}

A well-known feature of rapid granular flows is the instability of
the homogeneous cooling state against long wavelength spatial
perturbations, leading to cluster and vortex formation. Although the
origin of this instability has been widely explored by using computational tools,
prior analytical work on this subject has been limited to weak inelasticity or
very dilute regime. In order to gain some insight into the influence of both density and
dissipation on the stability of the HCS, a kinetic theory description has been adopted.
For moderate densities, the inelastic Enskog equation \cite{BDS97} can be considered as
a valuable tool for granular media. As in the case of elastic collisions, the Enskog equation takes
into account spatial correlations through the pair correlation function but neglects the
velocity correlations between the particles that are about to collide (molecular chaos).
The latter assumption has been clearly shown to fail for inelastic collisions as the density
increases \cite{MD}, so that the limitations of the Enskog description are greater than
for elastic collisions. Due to this molecular chaos breakdown, some authors conclude that
the Enskog equation can be insufficient to compute average properties of inelastic fluids,
except for very weak dissipation.
Nevertheless, this conclusion contrasts with previous comparisons made
with MD simulations
\cite{LBD02,DHGD02,AL03} and with real experiments \cite{YHCMW02} where, at least
for the problems studied there, velocity correlations do not seem to play an important role
and the Enskog
equation provides quite good estimates for the transport properties of the system.
It is remarkable that its accuracy is not restricted to the quasielastic limit since it
covers values of moderate density ($0\leq \phi \lesssim 0.15$) and large values of dissipation
($0\leq \alpha \lesssim 0.5$). It is possible that for situations more complex than those analyzed
in Refs.\ \cite{LBD02,DHGD02,AL03,YHCMW02,HYCMW04}, velocity correlations become important and the
Enskog theory does not give reliable predictions. In this case, new kinetic theories incorporating
the effect of velocity correlations are needed to describe granular flows. However, so far
there is no alternative to the Enskog theory for
finite density systems at this point. Hence, it is the most accurate
theory to describe systems of interest in simulations and experiments.

In this paper I have used the inelastic Enskog kinetic theory to perform a linear stability
analysis of the hydrodynamic equations and identify the conditions for stability in terms of
dissipation and density. The analysis is based on a previous derivation
\cite{GD99} of the expressions of the Enskog transport coefficients and the
cooling rate that, a priori, is not limited to small dissipation.
This is the main new ingredient of this work since previous studies
\cite{NE00} on linear stability analysis for dense granular gases considered weakly
inelastic systems and so, thermodynamic and
transport properties were assumed to be the same as those of {\em elastic} hard sphere
fluids. However, this assumption is expected to fail as dissipation increases since the form
of the {\em inelastic} transport coefficients clearly differs from their elastic counterparts, as
shown for instance in Fig.\ \ref{fig1rev}.

The study reported here extends to higher densities a previous linear stability analysis performed for
a dilute gas \cite{BDKS98}. In general, the findings agree qualitatively well with previous results
\cite{NE00}, showing that the effect of dissipation on transport coefficients do not significantly
modify the qualitative form of the dispersion relations. Specifically, linear stability analysis shows two
unstable modes: a transversal shear mode and a longitudinal ``heat'' mode. The instability of both modes is
a long wavelength instability.
The analysis of the dependence of the corresponding critical shear
wave number $k_s$ [defined in Eq.\ (\ref{16.3})]  and heat wave number
$k_h$ [defined in Eq.\ (\ref{29})] shows that,
except for extreme values of dissipation, the instability is
driven by the transversal shear mode. The range of values of the coefficient of restitution $\alpha$
for which $k_h>k_s$ is shortened as the gas becomes denser. Thus, for $\phi\gtrsim 0.389$, $k_s>k_h$
for any value of $\alpha$.

On the other hand, as expected, quantitative discrepancies between our results
and those given in Ref.\ \cite{NE00} become significant as the dissipation increases.
In particular, at
a given value of density, the critical wave numbers $k_s$ and $k_h$
are in general underestimated (except in the case of $k_s$ for a low-density gas)
when one neglects the influence of inelasticity on transport \cite{NE00} while
the ratio $k_s/k_h$ (which is independent of the coefficient of restitution $\alpha$ in \cite{NE00})
presents a complex dependence on the rate of dissipation, as is illustrated in Fig.\ \ref{fig6rev}.
Therefore, although the description made by van Noije and Ernst \cite{NE00} predicts reasonably well the
dispersion relations as well as the long-range structure, one expects that the results
reported here improve such predictions when one considers values of the coefficient
of restitution $\alpha$ for which transport properties are clearly affected by
the rate of dissipation.

As said in the Introduction, in the case of a dilute gas ($\phi=0$) comparison with direct
Monte Carlo simulation of the Boltzmann equation has shown the accuracy of the stability
analysis performed in Ref.\ \cite{BDKS98}. Given that the results reported here extends the above
description to high densities, comparison with MD simulations
becomes practical.
In this context, it is hoped that the description reported here stimulates the performance of
such computer simulations to characterize the onset and evolution of the clustering instability. As in the
Boltzmann case \cite{BRM98,BRD99}, one expects that the Enskog results describes accurately
the first stages of evolution.

Finally, it must noted that all the results obtained in this paper has been made in the context of
a very simple collision model where the coefficient of restitution is constant. Recent results \cite{BSSP04}
derived with an impact-velocity-dependent coefficient of restitution shows that structure formation
occurs in free granular gases only as a transient phenomenon, whose duration increases with the system size.

\acknowledgments
I am grateful to Dr. James Lutsko for pointing out some errors
in the evaluation of the terms coming from the cooling rate to first order in the gradients.
Partial support of the Ministerio de Educaci\'on y Ciencia (Spain) through Grant No. FIS2004-01399
(partially financed by FEDER funds) is acknowledged.

\appendix
\section{}
\label{appA}
In this Appendix, the expressions for the hydrostatic pressure, the transport coefficients
and the cooling rate used in Eqs.\ (\ref{4})--(\ref{8}) are given. The reduced hydrostatic
pressure $p^*$ is
\begin{equation}
\label{a1}
p^*=1+2(1+\alpha)\phi \chi
\end{equation}
The reduced transport coefficients $\eta^*$, $\gamma^*$, $\kappa^*$, and $\mu^*$ defined through
the relations (\ref{7}) and (\ref{8}) are given, respectively by
\begin{equation}
\label{a2}
\eta^*=\eta_k^*\left[1+\frac{4}{5}\phi \chi (1+\alpha)\right]+\frac{3}{5}\gamma^*,
\end{equation}
\begin{equation}
\label{a3}
\gamma^*=\frac{128}{5\pi}\phi^2 \chi (1+\alpha)\left(1-\frac{c}{32} \right),
\end{equation}
\begin{equation}
\label{a4}
\kappa^*=\kappa_k^*\left[1+\frac{6}{5}\phi \chi (1+\alpha)\right]+\frac{256}{25\pi}
\phi^2 \chi (1+\alpha)\left(1+\frac{7}{32} c \right),
\end{equation}
\begin{equation}
\label{a5}
\mu^*=\mu_k^*\left[1+\frac{6}{5}\phi \chi (1+\alpha)\right].
\end{equation}
Here, the superscript $k$ denotes the contributions to the transport coefficients coming
from the kinetic parts of the fluxes \cite{GD99}. These kinetic contributions are
\begin{equation}
\label{a6}
\eta_k^*=\left(\nu_{\eta}^*-\frac{1}{2}\zeta_0^*\right)^{-1}\left[1-\frac{2}{5}(1+\alpha)
(1-3 \alpha)\phi \chi \right],
\end{equation}
\begin{eqnarray}
\label{a7}
\kappa_k^*&=&\frac{2}{3}\left(\nu_{\kappa}^*-2\zeta_0^*\right)^{-1}
\nonumber\\
& & \times
\left\{1+\left[1+(1+\alpha)\phi \chi \right]c +\frac{3}{5}\phi \chi(1+\alpha)^2\right. \nonumber\\
& & \left.\times  \left[2 \alpha-1+\left( \frac{1+\alpha}{2}
-\frac{5}{3(1+\alpha)}\right)
c\right]\right\},
\end{eqnarray}
\begin{eqnarray}
\label{a8}
\mu_k^*&=&2\left(2\nu_{\kappa}^*-3\zeta_0^*\right)^{-1}\left\{\left(1+
\phi \partial_{\phi} \ln \chi\right) \zeta_0^*\kappa_k^*\right.
\nonumber\\
& &
+\frac{1}{3}p^*(1+\phi \partial_{\phi} \ln p^*)c
-\frac{4}{5}\phi \chi (1+\alpha)\nonumber\\
& & \times
\left(1+\frac{1}{2}\phi \partial_{\phi} \ln \chi \right)
\left[\alpha(1-\alpha) \right.
\nonumber\\
& & \left.\left.
+\frac{1}{4}\left(\frac{4}{3}+\alpha(1-\alpha)\right)c\right]\right\}.
\end{eqnarray}
In these expressions, $c$, $\zeta_0^*$, $\nu_{\eta}^*$, and $\nu_{\kappa}^*$ are functions of
$\alpha$ and $\phi$ given by
\begin{equation}
\label{a9}
c=\frac{32(1-\alpha)(1-2 \alpha^2)}{81-17 \alpha+30 \alpha^2(1-\alpha)},
\end{equation}
\begin{equation}
\label{10}
\zeta_0^*=\frac{5}{12}\chi (1-\alpha^2)\left(1+\frac{3}{32}c\right),
\end{equation}
\begin{equation}
\label{a11}
\nu_{\eta}^*=\chi \left[1-\frac{1}{4}(1-\alpha)^2\right]\left(1-\frac{c}{64}\right),
\end{equation}
\begin{equation}
\label{a12}
\nu_{\kappa}^*=\frac{1}{3}\chi (1+\alpha)\left[1+\frac{33}{16}(1-\alpha)+
\frac{19-3\alpha}{1024} c\right].
\end{equation}
Furthermore, in three dimensions the Carnahan-Starling approximation \cite{CS69} for the
the pair correlation function at contact $\chi(\phi)$ is given by
\begin{equation}
\label{a11bis}
\chi(\phi)=\frac{2-\phi}{2(1-\phi)^3}.
\end{equation}

The coefficient $\zeta_1$ appearing in the expression (\ref{6}) for the cooling rate
$\zeta$ is
\begin{equation}
\label{a13}
\zeta_1=\left[\frac{5}{32}\left(1+\frac{3}{64}c\right)c_{\zeta}-2\right]\phi \chi (1-\alpha^2),
\end{equation}
where \cite{footnote}
\begin{equation}
\label{a14}
c_{\zeta}=\frac{\frac{4}{15}\lambda+(1+\alpha)\left(\frac{1}{3}
-\alpha\right)c}{\nu_{\zeta}^*
-\frac{5}{8}(1-\alpha^2)\left(1+\frac{3}{32}c\right)+\frac{5c}{64}
\left(1+\frac{3}{64}c\right)(1-\alpha^2)},
\end{equation}
\begin{equation}
\label{a15}
\nu_{\zeta}^*=\frac{1+\alpha}{192}\left[241-177 \alpha+30 \alpha^2-30 \alpha^3
+\frac{c}{64}
\left(30 \alpha^3-30 \alpha^2+2001 \alpha-1873\right)\right]
\end{equation}
\begin{equation}
\label{a16}
\lambda=\frac{3}{8}(1+\alpha)\left[(1-\alpha)(5 \alpha^2+4 \alpha-1)
+\frac{c}{12}
\left(159 \alpha+3 \alpha^2-19 -15 \alpha^3\right)\right].
\end{equation}

For elastic collisions ($\alpha=1$), $c=\zeta_0^*=\mu^*=\zeta_1=0$ and
\begin{equation}
\label{a17}
p^*(1,\phi)=1+4\chi \phi , \quad
\gamma^*(1,\phi)=\frac{256}{5\pi}\chi \phi^2,
\end{equation}
\begin{equation}
\label{a18}
\eta^*(1,\phi)=\chi^{-1}\left(1+\frac{8}{15}\chi \phi\right)^2+\frac{3}{5}\gamma^*,
\end{equation}
\begin{equation}
\label{a19}
\kappa^*(1,\phi)=\chi^{-1}\left(1+\frac{12}{5}\chi \phi\right)^2+\frac{2}{5}\gamma^*.
\end{equation}
These expressions coincide with well-know results derived for normal hard-sphere fluids \cite{CC70}.

\end{document}